\documentclass[aps,pre,twocolumn,superscriptaddress,floatfix]{revtex4}
\usepackage{graphicx}
\usepackage[dvips,unicode,colorlinks,linkcolor=blue,citecolor=blue,urlcolor=blue]{hyperref}
\usepackage{times}
\begin{document}
\title{Localized defect modes in graphene}

\author{Alexander V. Savin}

\affiliation{Nonlinear Physics Center, Research School of Physics and
Engineering, Australian National University, Canberra ACT 0200,
Australia}

\affiliation{Semenov Institute of Chemical Physics, Russian Academy
of Sciences, Moscow 119991, Russia}

\author{Yuri S. Kivshar}
\affiliation{Nonlinear Physics Center, Research School of Physics and
Engineering, Australian National University, Canberra ACT 0200,
Australia}

\begin{abstract}
We study the properties of localized vibrational modes associated with structural defects in a
sheet of graphene. For the example of the Stone-Wales defects, one- and two-atom vacancies,
many-atom linear vacancies, and adatoms in a honeycomb lattice, we demonstrate that the local
defect modes are characterized by stable oscillations with the frequencies lying outside the linear
frequency bands of an ideal graphene. In the frequency spectral density of thermal oscillations,
such localized defect modes lead to the additional peaks from the right side of the frequency band
of the ideal sheet of graphene. Thus, the general structure of the frequency spectral density can
provide a fingerprint of its quality and the type of quantity of the structural defect a graphene
sheet may contain.
\end{abstract}

\maketitle

\section{Introduction\label{sc1}}

Over the past 20 years nanotechnology has made an impressive impact on the development of many fields
of physics, chemistry, medicine, and nanoscale engineering~\cite{book}. After the discovery of graphene
as a novel material for nanotechnology~\cite{science}, many properties of this two-dimensional object
have been studied both theoretically and experimentally~\cite{review,review2}.

Usually, depending on the type of growth and fabrication procedure employed, the graphene surface
has structural defects which may change substantially its mechanical, electronic, and transport properties.
Structural defects may also change both electronic and phononic spectra of the graphene. Moreover, defects
creates additional sources of scattering for phonons and electrons, so they may alter substantially
the transport properties of graphene structures.

By adding defects to an ideal graphene, we may change its properties, and this is
the main concept of nanoengineering of defect structures on graphene. In particular, local defects
may change the absorbing properties of graphene~\cite{ZZ2010}, whereas extended defects allow
the use of graphene as a membrane for gas separation~\cite{QM2013}, and arrays of defects may
change the conducting properties of graphene structures~\cite{AC2010,SL2012}.

We may divide various localized defects in graphene into two classes: point defects 
(such as Stone-Wales defect, vacancies, and adatoms) and one-dimensional dislocation-like defects~\cite{Banhart2011}.
Point defects in graphene act as scattering centers for electron and phonon waves
\cite{Gorjizadeh,Chen,Haskins2011}. Linear defects are responsible for plastic deformations of
graphene nanoribbons.

In this paper, we study the oscillatory phonon modes localized at the point defects in graphene.
To the best of our knowledge, this problem was never analyzed in detail. In particular, we apply
the molecular-dynamics numerical simulations and demonstrate that each type of structural defects
supports a number of localized modes with the frequencies located outside the frequency bands
of the ideal graphene lattice.  We also analyze nonlinear properties of these localized modes
and demonstrate that their frequencies decrease when the input power grows. These localized
defect modes manifest themselves in the thermalized dynamics of the graphene sheet,
and they may be employed as fingerprints of the graphene structural quality.

The paper is organized as follows. In Sec.~\ref{sc2} we describe our microscopic model and
introduce the interaction potentials. Section~\ref{sc3} summarizes the types of the local defects
in graphene analyzed in the paper. Sections \ref{sc4} and \ref{sc5} are devoted to the analysis
of the properties of in-plane and out-of-plane defect modes. In Sec.~\ref{sc6} and \ref{sc7}
we analyze how the defect modes are manifested in the dynamics and thermalized relaxation
of graphene structures. Section \ref{sc8} concludes the paper.

\section{Model\label{sc2}}

To model numerically the oscillations of the graphene sheet with local defects, we consider a
rectangular graphene lattice of a finite extent (a graphene flake) with the size
$10.8\times 10.6$ nm$^2$ composed of $N=4400$ carbon atoms. We place a defect at the center
of this graphene flake, and introduce periodic boundary condition to avoid the interaction
with the boundaries.

\begin{figure}[t]
\begin{center}
\includegraphics[angle=0, width=1\linewidth]{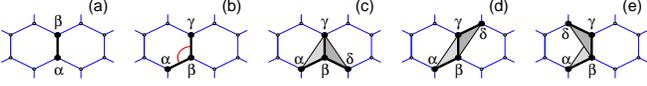}
\end{center}
\caption{\label{fg01}\protect
Configurations containing up to $i$th nearest-neighbor interactions
for (a) $i=1$, (b) $i=2$, (c) $i=3$, (d) $i=4$, and (e) $i=5$.
}
\end{figure}

To describe graphene oscillations, we write the system Hamiltonian in the form
\begin{equation}
H=\sum_{n=1}^N[\frac12M(\dot{\bf u}_n,\dot{\bf u}_n)+P_n],
\label{f1}
\end{equation}
where $M=12m_p$ is the mass of carbon atom ($m_p$ is the proton mass),
${\bf u}_n=[x_n(t),y_n(t),z_n(t)]$ is the radius vector
of the carbon atom with the index $n$ at the moment $t$.
The term $P_n$ describes the interaction of the atom with the index $n$
and its neighboring atoms. The potential energy depends on variations in bond length, bond angles,
and dihedral angles between the planes formed by three neighboring carbon atoms
and it can be written in the form
\begin{equation}
P=\sum_{\Omega_1}U_1+\sum_{\Omega_2}U_2+\sum_{\Omega_3}U_3+\sum_{\Omega_4}U_4+\sum_{\Omega_5}U_5,
\label{f2}
\end{equation}
where $\Omega_i$, with $i=1,2,3,4,5$, are the sets of configurations including up
to nearest-neighbor interactions. Owing to a large redundancy, the sets only need to contain
configurations of the atoms shown in Fig.~\ref{fg01}, including their rotated and mirrored versions.

The potential $U_1({\bf u}_\alpha,{\bf u}_\beta)$ describes the deformation energy
due to a direct interaction between pairs of atoms with the indexes
$\alpha$ and $\beta$, as shown in Fig. \ref{fg01}(a).
The potential $U_2({\bf u}_\alpha,{\bf u}_\beta,{\bf u}_\gamma)$
describes the deformation energy of the angle between the valent bonds
${\bf u}_\alpha{\bf u}_\beta$ and  ${\bf u}_\beta{\bf u}_\gamma$, see Fig.~\ref{fg01}(b).
Potentials $U_{i}({\bf u}_\alpha,{\bf u}_\beta,{\bf u}_\gamma,{\bf u}_\delta)$, $i=3$, 4, 5,
describes the deformation energy associated with a change of the effective angle between
the planes ${\bf u}_\alpha,{\bf u}_\beta,{\bf u}_\gamma$ and
${\bf u}_\beta,{\bf u}_\gamma,{\bf u}_\delta$, as shown in Figs. \ref{fg01}(c,d,e).

We use the potentials employed in the modeling of the dynamics of large polymer
macromolecules \cite{p7,p8,p9,p10,p11} for the valent bond coupling,
\begin{equation}
U_1({\bf u}_1,{\bf u}_2)=\epsilon_1
\{\exp[-\alpha_0(\rho-\rho_0)]-1\}^2,~\rho=|{\bf u}_2-{\bf u}_1|,
\label{f3}
\end{equation}
where $\epsilon_1=4.9632$~eV is the energy of the valent bond and $\rho_0=1.418$~\AA~
is the equilibrium length of the bond;
the potential of the valent angle
\begin{eqnarray}
U_2({\bf u}_1,{\bf u}_2,{\bf u}_3)=\epsilon_2(\cos\varphi-\cos\varphi_0)^2,
\label{f4}\\
\cos\varphi=({\bf u}_3-{\bf u}_2,{\bf u}_1-{\bf u}_2)/
(|{\bf u}_3-{\bf u}_2|\cdot |{\bf u}_2-{\bf u}_1|),
\nonumber
\end{eqnarray}
so that the equilibrium value of the angle is defined as $\cos\varphi_0=\cos(2\pi/3)=-1/2$;
the potential of the torsion angle
\begin{eqnarray}
\label{f5}
U_i({\bf u}_1,{\bf u}_2,{\bf u}_3,{\bf u}_4)=\epsilon_i(1-z_i\cos\phi),\\
\cos\phi=({\bf v}_1,{\bf v}_2)/(|{\bf v}_1|\cdot |{\bf v}_2|),\nonumber \\
{\bf v}_1=({\bf u}_2-{\bf u}_1)\times ({\bf u}_3-{\bf u}_2), \nonumber \\
{\bf v}_2=({\bf u}_3-{\bf u}_2)\times ({\bf u}_3-{\bf u}_4), \nonumber
\end{eqnarray}
where the sign $z_i=1$ for the indices $i=3,4$ (equilibrium value of the torsional angle $\phi_0=0$)
and $z_i=-1$ for the index $i=5$ ($\phi_0=\pi$).

The specific values of the parameters are $\alpha_0=1.7889$~\AA$^{-1}$,
$\epsilon_2=1.3143$ eV, and $\epsilon_3=0.499$ eV, and they are found from the frequency
spectrum of small-amplitude oscillations of a sheet of graphite~\cite{p12}.
According to the results of Ref.~\cite{p13} the energy $\epsilon_4$ is close to
the energy $\epsilon_3$, whereas  $\epsilon_5\ll \epsilon_4$
($|\epsilon_5/\epsilon_4|<1/20$). Therefore, in what follows we use the values
$\epsilon_4=\epsilon_3=0.499$ eV and assume $\epsilon_5=0$, the latter means that we
omit the last term in the sum (\ref{f2}).

More detailed discussion and motivation of our choice of the interaction potentials
(\ref{f2}), (\ref{f3}), (\ref{f4}) can be found in Ref.~\cite{skh10}. Such potentials have been employed
for modeling of thermal conductivity of carbon nanotubes~\cite{skh09,shk09}
graphene nanoribbons~\cite{skh10} and also in the analysis of their oscillatory modes~\cite{sk09,sk10,sk10prb}.\

\section{Types of localized defects in graphene\label{sc3}}

First, we introduce a model of graphene with defects. We take a rectangular flake of graphene
with an ideal honeycomb lattice and, depending on the type of defect we wish to create, we add or
remove some carbon atoms at its center by cutting the corresponding bonds. In this way, we define
the atomic configuration $\{ {\bf u}_n\}_{n=1}^N$ that will relax to the structure with a specific
type of defect.  To find the ground state of the graphene layer with defects, we should find
the energy minimum for the interaction energy,
\begin{equation}
\sum_{n=1}^N P_n \rightarrow\min: \{ {\bf u_n}\}_{n=1}^N.
\label{f6}
\end{equation}

Problem (\ref{f6}) is solved numerically by means of the conjugate gradient method. If $\{ {\bf u}_n^0\}_{n=1}^N$
is the ground state of this rectangular flake of graphene, then for small-amplitude oscillations
we can write ${\bf u}_n(t)={\bf u}_n^0+{\bf v}_n(t)$, where
$|{\bf v}_n|\ll \rho_0$. Then, the equations of motion corresponding to the Hamiltonian (\ref{f1})
can be written as a system of $3N$ linear equation for $3N$ variables,
\begin{equation}
-M\ddot{\bf v}_n=\frac{\partial H}{\partial{\bf u}_n}=\sum_{j=1}^NB_{jn}{\bf v}_j,~~
B_{jn}=\left.\frac{\partial^2 H}{\partial {\bf u}_n\partial{\bf u}_j}\right|_{\{ {\bf u}^0_k\}_{k=1}^N}.
\label{f7}
\end{equation}

To find all linear modes of the graphene sheet, we need to find numerically all
$3N\times3N$ eigenvalues and corresponding eigenvectors of the real symmetric
matrix ${\bf B}=(B_{jn})_{j,n=1}^N$. If we define $\lambda$ and
${\bf e}=\{{\bf v}^0_n\}_{n=1}^N$ as the eigenvalue and normalized eigenvector, namely
${\bf B}{\bf e}=\lambda {\bf e}$ and $({\bf e},{\bf e})=\sum_n({\bf v}^0_n,{\bf v}^0_n)=1$, then
the solution of Eq.~(\ref{f7}) will have the form
${\bf v}_n(t)=A{\bf v}_n^0\exp(i\omega t)$, where
$\omega=\sqrt{\lambda/M}$) is the mode frequency and $A$ is the mode amplitude.

We characterize the degree of spatial localization of any of the oscillatory eigenmode
by the parameter of localization (inverse participation number),
$d=\sum_{n=1}^N({\bf v}_n^0,{\bf v}_n^0)^2$.  For the modes
which are not localized in space, $d\approx 1/N$, for the mode localized on a
single atom, $d=1$. Inverse value $N_d=1/d$ (participation number) characterizes
the number of atoms which are involved into this oscillatory mode.
\begin{figure}[t]
\begin{center}
\includegraphics[angle=0, width=1\linewidth]{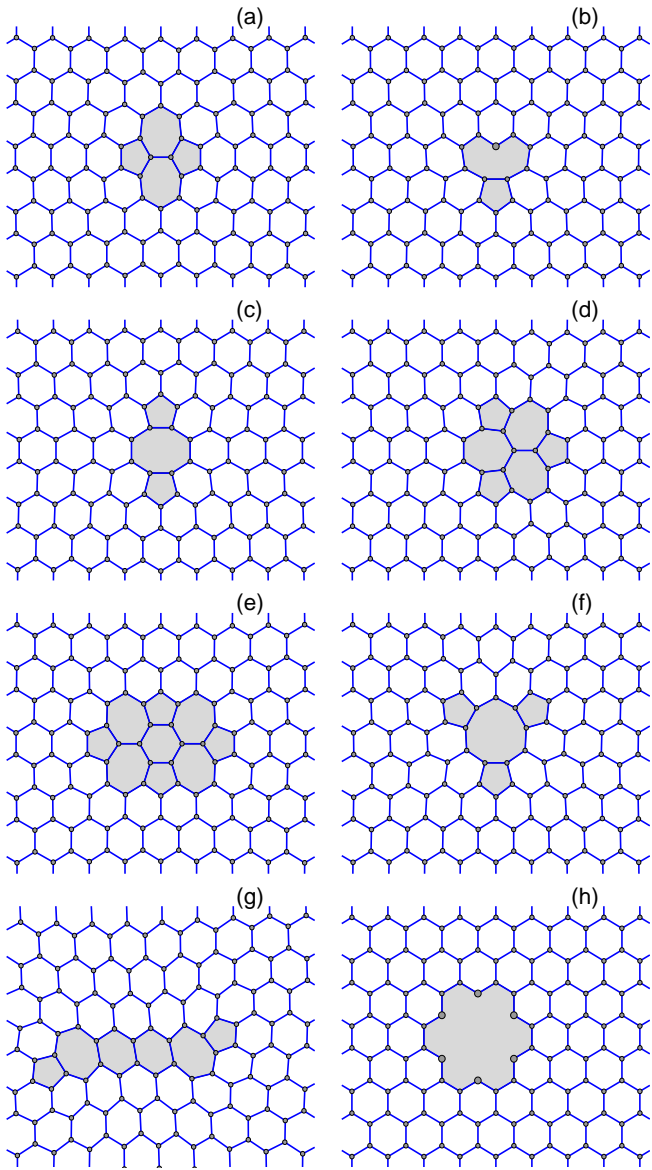}
\end{center}
\caption{\label{fg02}\protect
Localized defects which do not change the flat surface of the graphene layer,
(a) Stone-Wales defect SW(55-77),
(b) single vacancy V$_1$(5-9);
(c) double vacancy V$_2$(5-8-5);
(d) double vacancy V$_2$(555-777);
(e) double vacancy V$_2$(5555-6-7777);
(f) quadruple vacancy V$_4$(555-9);
(g) vacancy V$_8$(5-7-66-7-5),
extended linear defect obtained by removing a zigzag-like array of 8 carbon atoms;
(h) vacancy V$_6$, a defect created by removing 6 carbon atoms
with the formation of a hexagonal hole.   Grey color is employed to show
a change of the lattice structure introduced by the presence of a defect.
}
\end{figure}

For an ideal flake of graphene, all eigenmodes are delocalized and for the periodic boundary
conditions they are distributed homogeneously on the rectangular surface. All such oscillatory
modes can be divided into two classes: (i) in-plane modes, when the atoms oscillate along the flat
surface of the graphene sheet, and (ii) out-of-plane modes, when the atoms oscillate perpendicular
to the surface of the graphene sheet. The frequencies of the former modes are located
in the domain $0\le\omega\le 1600$~cm$^{-1}$, whereas the frequencies of the latter modes
correspond to a narrow spectral range,  $0\le\omega\le 900$~cm$^{-1}$.
\begin{figure}[t]
\begin{center}
\includegraphics[angle=0, width=1\linewidth]{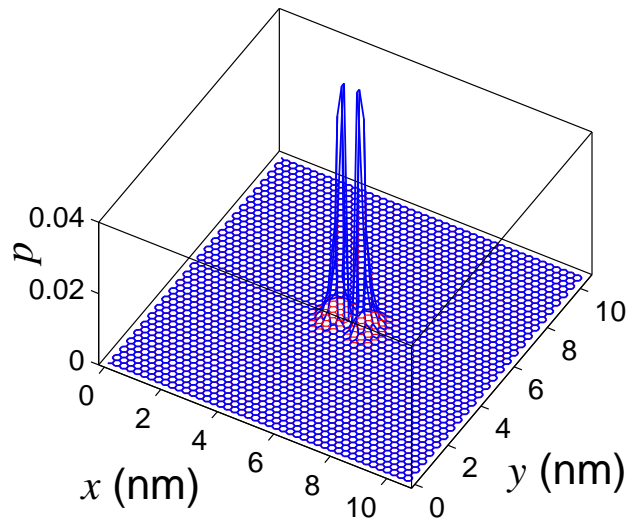}
\end{center}
\caption{\label{fg03}\protect
Energy of in-plane oscillations for the Stone-Wales defect SW(55-77)
with the frequency $\omega_d=1613.2$ cm$^{-1}$, and the participation number $N_d=26.31$.
}
\end{figure}

Bellow, first we study the in-plane localized defects shown in Fig.~\ref{fg02} which do not
bend the surface of the graphene sheet (Sec.~\ref{sc4}), but then consider the case
of out-of-plane defects (Sec.~\ref{sc5}).

\section{Oscillations of in-plane defects\label{sc4}}

To create the Stone-Wales defect SW(55-77), we should rotate one valent bond by 90 degrees.
Then four old neighboring valent bonds break creating four new bonds. As a result, we create
two pentagons and two heptagons (55-77) instead of four hexagons of the perfect honeycomb lattice,
as shown in Fig.~\ref{fg02}(a). In the ground state, this defect has the energy $E_d=2.06$~eV,
and it is stable. Although the defect's energy is higher than that of the ground state
by the value $E_d$, the reverse change of the bonds requires to overtake the energy barrier
of $\Delta E=9$~eV \cite{Banhart2011}.
\begin{figure}[t]
\begin{center}
\includegraphics[angle=0, width=1\linewidth]{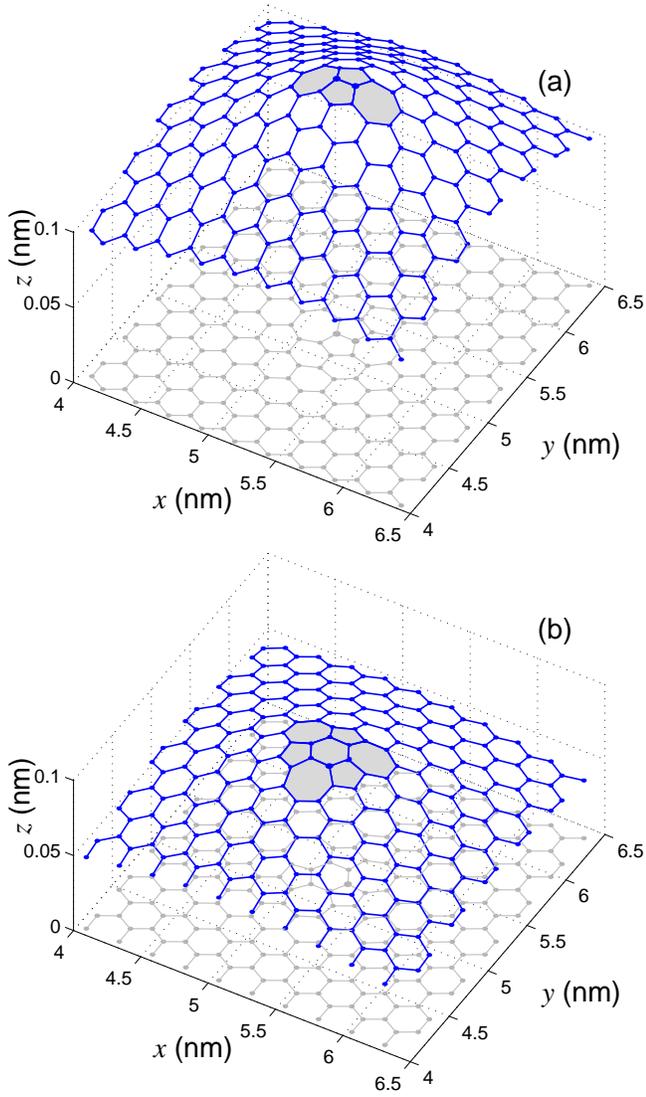}
\end{center}
\caption{\label{fg04}\protect
Deformation of the graphene sheet in the region of location of (a)
inverse Stone-Wales defect I$_2$(7557) (two carbon adatoms with a valent bond
are inserted in a perfect graphene lattice; they are shown by larger balls)
and (b) two adatoms defect A$_2$(5556777). Grey color is employed to show
the change of the lattice structure introduced by the presence of a defect.
}
\end{figure}

Our analysis of the localized modes of the lattice with the defect shows that the Stone-Wales
defect supports four in-plane localized modes with the frequencies
$\omega_d=1613.2$, 1614.1, 1615.6, 1616.3~cm$^{-1}$
(and the participation numbers $N_d=26.31$,  27.06,  25.65,  26.14)
and also four out-of-plane localized modes with the frequencies
$\omega_d=904.7$, 907.5, 909.4, 935.1~cm$^{-1}$ ($N_d=44.74$, 21.81, 37.24, 7.90).
The characteristic profile of the oscillatory energy of this defect mode is shown in Fig.~\ref{fg03},
where we observe that the energy is localized entirely in the region of defect, so the used
periodic boundary conditions do not have any influence on the mode's frequency and shape.
The oscillatory dynamics of all localized modes is shown in movies from Supplementary Material.

In order to create a single vacancy V$_1$(5-9) in the graphene lattice, we remove just one carbon
atom breaking three valent bonds. The remaining lattice relax with two neighboring atoms linked
by a new valent bond, and instead of three hexagons the structure (5-9) is formed with one
pentagon and one nonagon, see Fig.~\ref{fg02}(b). In this defect structure, one carbon atom
has only two valent bonds. In order to keep the interaction potentials those bonds unchanged
in our calculations, we assume that this atom is coupled to a hydrogen atom and has mass
$M=13m_p$. Them the energy of the defect V$_1$(5-9) is found to be $E_d=2.26$~eV.
This defect supports three in-plane localized eigenmodes with the frequencies
$\omega_d=1600.6$, 1601.3, 1606.3~cm$^{-1}$ ($N_d=146.99$, 79.41, 14.58) and two
out-of-plane lightly localized modes with the frequencies
$\omega_d=1227.0$, 1791.0~cm$^{-1}$ ($N_d=5.03$,  2.65).

Double-vacancy defect  V$_2$(5-8-5) shown in Fig.~\ref{fg02}(c) can be created by removing two
neighboring carbon atoms, breaking four valent bonds. After this procedure, the remaining four
neighboring carbon atoms create two new valent bonds, and in the lattice structure four hexagons
are replaced by two pentagons and one octagon. The energy of the double-vacancy defect
V$_2$(5-8-5) is found to be $E_d=3.44$~eV.
This defect supports four in-plane eigenmodes of localized oscillations with the frequencies
$\omega_d=1605.9$, 1606.0, 1619.2, 1619.3~cm$^{-1}$ ($N_d=65.91$,  66.16,  15.21,  15.32) and four
out-of-plane localized modes with the frequencies $\omega_d=902.6$,  905.0,  986.4, 1015.0~cm$^{-1}$
($N_d=64.16$,  57.77,   6.18,   7.27).

In order to create  the double-vacancy defect V$_2$(555-777),  we should start from the defect
V$_2$ (5-8-5) and rotate one of the bonds of its octagon by 90 degrees. Then four old bonds break
and four new bonds appear. In the graphene lattice we obtain instead of seven hexagons three pentagons
and three heptagons (555-777), as shown in Fig.~\ref{fg02}(d). This defect is characterized by the energy
$E_d=2.91$~eV.  Our analysis of the eigenmodes localized on this defect demonstrate that it has
no in-plane localized modes at all, but it supports four out-of-plane modes with the frequencies
$\omega_d=909.1$,  916.0,  916.0,  916.6~cm$^{-1}$ ($N_d=46.22$,  15.87,  15.87,  11.30).
\begin{figure}[t]
\begin{center}
\includegraphics[angle=0, width=1\linewidth]{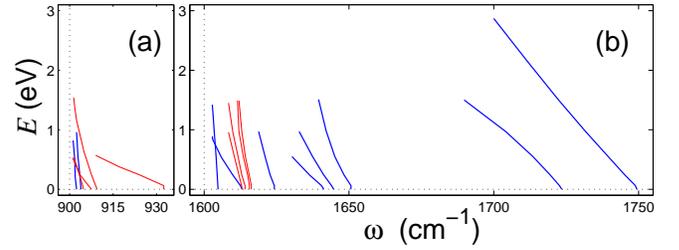}
\end{center}
\caption{\label{fg06}\protect
Dependencies of the energy $E$ of localized defect modes vs. frequency $\omega$,
for two types of defects: inverse Stone-Wales defect I$_2$(7557), shown by blue lines,
and for the Stone-Wales defect SW(55-77), shown by red lines. Parts (a) and (b) show
out-of-plane and in-plane oscillations, respectively. Vertical dotted line marks
the edge of the linear spectrum band of phonons of the ideal graphene sheet.
}
\end{figure}
\begin{figure}[t]
\begin{center}
\includegraphics[angle=0, width=1\linewidth]{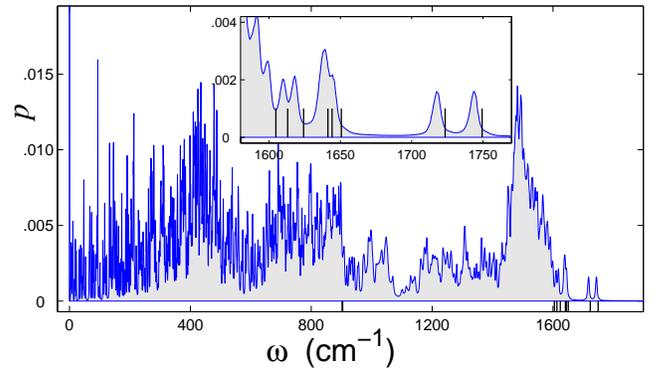}
\end{center}
\caption{\label{fg05}\protect
Frequency spectral density of thermal oscillations of a graphene sheet
with inverse Stone-Wales defect I$_2$, when two adatoms are inserted
into the honeycomb lattice. The graphene sample consists of $N=578$ atoms
and it has the size of $3.93\times 3.83$ nm$^2$, being placed at temperature of
 $T=300$K. Vertical solid bars show the frequencies of the localized oscillation
 of the defect modes.
}
\end{figure}

Double-vacancy defect V$_2$(5555-6-7777) is formed from the defect V$_2$(555-777) by another bond rotation.
In the honeycomb lattice, instead of ten perfect hexagons we have four pentagons, one hexagons,
and four heptagons, as shown in Fig.~\ref{fg02}(e). The energy of this double-vacancy defect is
$E_d=2.96$~eV. This defect supports seven localized in-plane eigenmodes with the frequencies
$\omega_d=1605.5$, 1605.6, 1607.3, 1607.4, 1666.4, 1773.7, 2009.4~cm$^{-1}$
(and the participation numbers $N_d=25.84$, 25.50, 25.99, 23.65, 5.86, 4.57, 3.05)
and four out-of-plane localized modes with the frequencies
$\omega_d=905.7$, 907.2, 916.4, 918.4~cm$^{-1}$ ($N_d=46.72$,  54.22,  15.07,  14.52).

The largest hole appears when we remove 4 neighboring carbon atoms from the graphene lattice
with {\textsf Y}-shape valent bonds.
After that, six carbon atoms coupled to the removed atom got involved
into new valent bonds, thus creating quadruple vacancy defect V$_4$(555-9). In this defect,
instead of six hexagons we have three heptagons and one large nonagon, as shown in Fig.~\ref{fg02}(f).
The energy of this defect is $E_d=5.37$~eV.  This defect supports three in-plane eigenmodes
with the frequencies $\omega_d=1601.9$, 1602.4, 1602.4~cm$^{-1}$
(participation numbers $N_d=59.02$, 40.46, 40.21) and also four out-of-plane modes
with the frequencies $\omega_d=902.3$, 908.9, 976.6, 997.3~cm$^{-1}$ ($N_d=52.27$, 11.01, 7.23, 6.51).

For comparison, next we consider an extended defect in the form of the vacancy V$_8$(5-7-66-7-5),
that appears after removing a zigzag chain of an array of eight carbon atoms shown in Fig.~\ref{fg02}(g).
In the honeycomb lattice, instead of ten perfect hexagons we create two pentagons, two hexagons,
and two heptagons; the defect energy is $E_d=11.92$~eV. This extended defect supports
seven in-plane localized modes with the frequencies
$\omega_d=1604.2$, 1606.8, 1607.2, 1613.4, 1624.3, 1627.3, 1653.2~cm$^{-1}$
($N_d=57.72$,  19.39,  18.03,  11.53,  10.67,  11.49,   4.36) and also four out-of-plane
localized modes with the frequencies $\omega_d=909.3$,  910.5,  927.2, 1091.0~cm$^{-1}$
corresponding to $N_d=21.35$,  21.53,   6.25, and  2.69.

If we remove one hexagon from the honeycomb lattice, we create a vacancy V$_6$ that can not be "healed"
\ by the lattice transformations. This vacancy has the form of a perfect hexagon hole with the diameter
of 0.57 nm, with almost unchanged old valent bonds, see Fig.~\ref{fg02}(h).  Our analysis demonstrates
that such holes do not support any localized modes.

\section{Oscillations of Out-of-plane defects\label{sc5}}

Unlike vacancies, adding extra atoms to the graphene sheet lead to a deformation of the planar structure
of the lattice. In the place where we have more atoms, the graphene sheet is no longer flat, and
it becomes either concave or convex, transforming into one of the two equivalent non-planar states.

When two migrating adatoms meet each other and form a dimer, they can be incorporated into
the graphene lattice, creating the so-called inverse Stone-Wales defect I$_2$(7557).
In such a state, instead of three hexagons we have two pentagons and two heptagons,
as shown in Fig.~\ref{fg04}(a). The surface of the sheet become deformed bending into either side.

Our analysis shows that such a defect supports 8 localized modes for which atoms move along
the curved surface. The mode eigenfrequencies are:
$\omega_d=1604.8$, 1613.1, 1624.2, 1641.3, 1644.2, 1650.7, 1723.5, 1749.4~cm$^{-1}$
(with the participation numbers $N_d=73.63$, 5.93,  19.82,  6.73, 8.64,  12.37,  3.14,  2.85).
This defect also supports two out-of-plane localized modes for which atoms move perpendicular
to the surface, the modes have the frequencies $\omega_d=902.4$, 903.9~cm$^{-1}$ ($N_d=99.55$, 71.79).

An excess of two carbon atoms may lead to another concave deformation of the graphene lattice
and two adatom defect mode A$_2$(5556777), when instead of 6 perfect hexagons we have 3 pentagons,
1 hexagon, and 3 heptagons, see Fig.~\ref{fg04}(b). Such a defect may appear when we
remove 4 carbon atoms from the graphene lattice with {\textsf Y}-shape valent bonds
but insert instead of them 6 carbon atoms with a structure of a perfect hexagon.
The transverse oscillations of this defect blister is considered in Ref.~\cite{PL2012}.
As can be seen from Fig.~\ref{fg04}, the defect I$_2$ has a higher density of the atom
localization in comparison with the defect A$_2$, and as a result the defect I$_2$
will bend the graphene sheet stronger than the defect A$_2$.

Our analysis of the localized oscillations of the graphene sheet demonstrate that the defect
A$_2$(5556777) supports 11 localized modes for which the atoms move along a concave surface of
the graphene sheet. The frequencies of those oscillations are:
$\omega_d=1601.0$, 1603.3, 1604.2, 1608.6, 1610.7, 1613.6, 1614.6, 1617.8, 1682.4, 1704.2, 1706.4~cm$^{-1}$
(participation numbers $N_d=131.01$, 19.52, 22.49, 13.37, 7.28, 7.96, 19.94, 4.48, 3.94, 3.52, 2.77).
This defect supports also two out-of-plane oscillatory modes for which atoms move perpendicular
to the graphene sheet; these modes have the frequencies: $\omega_d=900.7$, 910.4~cm$^{-1}$
($N_d=85.54$, 6.15).

\section{Modeling of localized oscillations\label{sc6}}

To verify the results of our modal analysis, we model the oscillatory dynamics of a graphene with
defects. We consider a graphene sheet of a finite extent, with the size of $10.8\times 10.6$~nm$^2$
with a local defect at its center. To model the energy spreading in an effectively infinite graphene
sheet, we introduce lossy boundary conditions assuming that all edge atoms experience damping with
relaxation time $t_r=10$~ps. If the excitation of the localized eigenmode lead to the creation of a
localized state, or breathing-like mode, the kinetic energy not vanish but instead it
will approach a certain nonzero value. Otherwise, the initial excitation will vanish completely,
so that the kinetic energy
$$
E_k=\sum_{n=1}^N\frac12 M(\dot{\bf u}_n,\dot{\bf u}_n)
$$
will vanish when $t\rightarrow\infty$.

Our numerical simulation results demonstrate that all localized modes with the frequencies located
outside the linear frequency bands of the ideal graphene sheet are stable. For small amplitudes,
when the energy of the eigenmode less than  0.06~eV, the amplitude and frequency of the mode
do not change, and this result indicates that in that limit the oscillations are described
by a linear theory. However, when more energy is pumped into the mode, we observe nonlinear
effects when the frequency decreases with the energy, as shown in Fig.~\ref{fg06}.
In that case, the energy of the defect mode may exceed 1~eV.

We notice that the diagonalization of the matrix  of the second derivatives ${\bf B}=(B_{jn})_{j,n=1}^N$
also can give localized modes with the frequencies within the frequency bands of
the ideal graphene sheet. However, numerical simulations demonstrate that such modes
are not stable and the amplitude decays very rapidly.

\section{Influence of defects on the graphene's frequency spectra\label{sc7} }

Our results summarized above suggest that all frequencies of localized defect modes are
located outside the linear bands of the oscillation spectrum of an ideal graphene sheet.
This result is expected, because any oscillation from the continuous spectrum
is not spatially localized, and it corresponds to one of the extended modes.
The splitting of the continuous modes of a perfect sheet of graphene, and localized modes
of the graphene with defects can be employed as an important fingerprint of the quality
of the graphene honeycomb lattice. We demonstrate this approach for the case of a graphene
sheet of a finite extent of $3.93\times 3.83$~nm$^2$ with inverse Stone-Wales defect I$_2$.
This patch of graphene has 578 carbon atoms two of which create one point-like defect
of the type shown in Fig.~\ref{fg04}(a).

In order to find the frequency spectrum of thermal oscillations of a graphene sheet, we place
it in a Langevin thermostat at temperature $T=300$~K. After complete thermalization,
we disconnect the graphene from the thermostat and study the oscillation spectra
of the excited thermal oscillations.

Frequency spectral density of thermal oscillations of a sheet of graphene is shown in Fig.~\ref{fg05}.
As follows from those results, several additional peaks appear in the frequency spectrum above the upper
edge of the spectral band of linear oscillations of a perfect graphene sheet ($\omega>1600$~cm$^{-1}$).
These peaks correspond to the eigenmodes of the localized oscillations of the defect I$_2$.
Thus, from the specific structure of the frequency spectral density we may judge about
 the types and density of structural defects in a particular graphene sample.

\section{Conclusions\label{sc8}}

We have studied the effect of localized structural defects on the vibrational spectra of the
graphene's oscillations. We have demonstrated that the local defects (such as Stone-Wales defects,
one- and two-atom vacancies, many-atom linear vacancies and adatoms) are characterized by stable
localized oscillations with the frequencies lying outside the linear frequency bands of the ideal graphene.
In the frequency spectral density of thermal oscillations, such localized defect modes lead 
to the additional peaks located on the right side from the frequency band of the ideal sheet 
of graphene. Thus, the general
structure of the frequency spectral density can provide a fingerprint of its quality
and the type of quantity of the structural defect a graphene sheet may contain.

\section*{Acknowledgements}

Alex Savin acknowledges a warm hospitality of the Nonlinear Physics Center at the Australian
National University, and thanks the Joint Supercomputer Center of the Russian Academy
of Sciences for the use of their computer facilities.
The work was supported by the Australian Research Council.


\begin{references}

\bibitem{book} K.E. Drexler, {\em Nanosystems: Molecular Machinery, Manufacturing,
and Computation} (New York, Wiley, 1992).

\bibitem{science}
K.S. Novoselov, A.K. Geim, S.V. Morozov, D. Jiang, Y. Zhang, S.V. Dubonos, I.V. Grigorieva,
and A.A. Firsov, Science {\bf 306}, 666 (2004).

\bibitem{review}
A.K. Geim and A.H. MacDonald, Phys. Today {\bf 60}, 35 (2007).

\bibitem{review2}
A.H. Castro Neto, F. Guinea, N.M. Peres, K.S. Novoselov, and A.K. Geim, 
Rev. Mod. Phys. {\bf 81}, 109 (2009).

\bibitem{ZZ2010}
Y.-H. Zhang, K.-G. Zhou, K.-F. Xie, X.-C. Gou, J. Zeng, H.-L. Zhang, and Y. Peng,
J. Nanosci. Nanotechnology, {\bf 10}, 7347 (2010).

\bibitem{QM2013}
X. Qin, Q. Meng, Y. Feng, and Y. Gao, Surface Science {\bf 607}, 153 (2013).

\bibitem{AC2010}
D.J. Appelhans, L.D. Carr and M.T. Lusk,
New J. Phys. {\bf 12}, 125006 (2010).

\bibitem{SL2012}
J. Song, H. Liu, H. Jiang, Q. Sun, and X. C. Xie,
Phys. Rev B {\bf 86}, 085437 (2012).

\bibitem{Banhart2011}
F. Banhart, J. Kotakoski, and A. V. Krasheninnikov,
ACS Nano {\bf 5}, 26 (2011).

\bibitem{Gorjizadeh}
N. Gorjizadeh, A. A. Farajian and Y. Kawazoe,
Nanotechnology {\bf 20}, 015201 (2009).

\bibitem{Chen}
J.-H. Chen, W. G. Cullen, C. Jang, M. S. Fuhrer, and E. D. Williams,
Phys. Rev. Lett. {\bf 102}, 236805 (2009).

\bibitem{Haskins2011}
J. Haskins, A. Kinaci, C. Sevik, H. Sevincli, G. Cuniberti, and T. Cagin
ACS Nano {\bf 5}, 3779 (2011).

\bibitem{p7} D. W. Noid, B. G. Sumpter, and B. Wunderlich, Macromolecules {\bf 24}, 4148 (1991).

\bibitem{p8} B. G. Sumpter, D. W. Noid, G. L. Liang, and B. Wunderlich, 
Adv. Polym. Sci. {\bf 116}, 27 (1994).

\bibitem{p9} A.V. Savin and L.I. Manevitch, Phys. Rev. B {\bf 58}, 11386 (1998).

\bibitem{p10} A.V. Savin and L.I. Manevitch, Phys. Rev. E {\bf 61}, 7065 (2000).

\bibitem{p11} A. V. Savin and L. I. Manevitch, Phys. Rev. B {\bf 67}, 144302 (2003).

\bibitem{p12} A. V. Savin and Yu. S. Kivshar, Europhys. Letters {\bf 82}, 66002 (2008).

\bibitem{p13} D. Gunlycke, H. M. Lawler, and C. T. White, Phys. Rev. B {\bf 77}, 014303 (2008).

\bibitem{skh10}
A. V. Savin, Yu. S. Kivshar,  and B. Hu.
Phys. Rev. B {\bf 82}, 195422 (2010).

\bibitem{skh09}
A. V. Savin, Yu. S. Kivshar, and B. Hu.
Europhys. Lett. {\bf 88}, 26004 (2009).

\bibitem{shk09}
A.V. Savin, B. Hu, and Yu. S. Kivshar.
Phys. Rev. B {\bf 80}, 195423 (2009).

\bibitem{sk09}
A. V. Savin and Yu. S. Kivshar.
Appl. Phys. Lett. {\bf 94}, 111903 (2009).

\bibitem{sk10}
A. V. Savin and Yu. S. Kivshar.
Europhys. Lett. {\bf 89}, 46001 (2010).

\bibitem{sk10prb}
A. V. Savin and Yu. S. Kivshar.
Phys. Rev. B {\bf 81}, 165418 (2010).

\bibitem{PL2012}
C.-W. Pao, T.-H. Liu, C.-C. Chang, and  D.J. Srolovitz,
Carbon {\bf 50}, 2870 (2012).

\end{references}
\end{document}